\documentclass[letterpaper,12pt]{article}
\usepackage{graphicx} 
\usepackage[final]{hyperref} 
\usepackage{physics}
\usepackage{authblk}
\hypersetup{
	colorlinks=true,       
	linkcolor=blue,        
	citecolor=red,        
	filecolor=magenta, 
	urlcolor=blue         
}

\usepackage{tikz}
\usetikzlibrary{arrows.meta,arrows}
\usepackage{amsmath,amssymb,amsfonts,amsthm} 
\usepackage[capitalize,nameinlink,noabbrev,nosort]{cleveref}

\newcommand{\ds}{\displaystyle}
\newcommand{\qbinom}[2]{\bgroup\renewcommand*{\arraystretch}{1}\begin{bmatrix} #1 \\ #2\end{bmatrix} \egroup}
\newcommand{\fracpart}[1]{\left\{ #1 \right\}}
\newcommand{\cZ}{\mathcal{Z}}
\newcommand{\vac}{\circ}
\newcommand{\occ}{\bullet}

\let\deg\relax
\DeclareMathOperator{\deg}{deg}
\DeclareMathOperator{\ord}{ord}
\DeclareMathOperator{\wt}{wt}
\DeclareMathOperator{\coinv}{coinv}

\begin{document}

\title{An exactly solvable asymmetric $K$-exclusion process}
\author[$\dagger$]{Arvind Ayyer}
\author[$\star$]{Samarth Misra}
\affil[$\dagger$]{Department of Mathematics, 
Indian Institute of Science, Bangalore  560012, India.}
\affil[$\star$]{Department of Physics, 
Indian Institute of Science, Bangalore  560012, India.}
\date{\today}

\maketitle

\begin{abstract}
We study an interacting particle process on a finite ring with $L$ sites with at most $K$ particles per site, in which particles hop to nearest neighbors with rates given in terms of $t$-deformed integers and asymmetry parameter $q$, where $t>0$ and $q \geq 0$ are parameters. This model, which we call the $(q, t)$~$K$-ASEP, reduces to the usual ASEP on the ring when $K = 1$ and to a model studied by Sch\"utz and Sandow (\emph{Phys. Rev. E}, 1994) when $t = q = 1$.
{This is a special case of the misanthrope process and as a consequence, the steady state does not depend on $q$ and is of product form, generalizing the same phenomena for the ASEP. }
{What is interesting here is the steady state weights are given by explicit formulas involving $t$-binomial coefficients, and are palindromic polynomials in $t$.}
Interestingly, although the $(q, t)$~$K$-ASEP does not satisfy particle-hole symmetry, its steady state does. We analyze the density and calculate the most probable number of particles at a site in the steady state in various regimes of $t$. Lastly, we construct a two-dimensional exclusion process on a discrete cylinder with height $K$ and circumference $L$ which projects to the $(q, t)$~$K$-ASEP and whose steady state distribution is also of product form. {We believe this model will serve as an illustrative example in constructing two-dimensional analogues of misanthrope processes.}

Simulations are attached as ancillary files.
\end{abstract}

\section{Introduction}

The asymmetric simple exclusion process (ASEP) is an important interacting particle system on both the finite and infinite one-dimensional integer lattice in which every site has at most one particle and particles hop to the neighboring site on the right (resp. left) with rate $1$ (resp. $q$) provided the target site is empty. Originally studied in the biophysics literature~\cite{macdonald-gibbs-pipkin-1968}, it was formulated rigorously by Spitzer~\cite{spitzer-1970} on the infinite integer lattice. The finite variant of the ASEP with open boundary conditions was solved exactly in~\cite{dehp}. 
Since then, the ASEP and its variants have proved to be invaluable testing grounds for formulating new principles for nonequilibrium statistical physics~\cite{bodineau-derrida-2004,BDJGL-2005}; see~\cite{Der07} for a review. The ASEP has also found connections to other areas in mathematics~\cite{corteel-williams-2011,cantini-etal-2016,corteel-et-al-2022}.

Many generalizations of the ASEP have been considered in the literature. A natural variant of the ASEP is the so-called \emph{$K$-exclusion process} (also called the \emph{partial exclusion process}), where we allow up to $K$ particles per site. The natural choice of rates is analogous to that of the ASEP, namely one particle at a site hops to the neighboring site on the right (resp. left) with rate $1$ (resp. $q$) provided the target site has less than $K$ particles. The symmetric ($q = 1$) version of this process on $\mathbb{Z}$ has been studied first by Keisling~\cite{keisling-1998} and the totally asymmetric version ($q = 0$) variant, again on $\mathbb{Z}$ first, by Sepp\"al\"ainen~\cite{seppalainen-1999}.

There is, however, another choice of rates which is also natural and which was first studied in the symmetric case by Sch\"utz--Sandow~\cite{schutz-sandow-1994}.
In their model, if a site has $a$ particles and its neighbor on either the right or left has $b$ particles, the rate of a particle moving from the former to the latter is $a(K-b)$. This is also natural because this is automatically $0$ if either $a = 0$ or $b = K$. Note also that this generalizes the symmetric exclusion process because the only nonzero transition occurs when $a = 1$ and $b = 0$ and this is $1$.
{A far-reaching generalisation of the Sch\"utz--Sandow model is the \emph{misanthrope process}, first introduced on $\mathbb{Z}$ by Cocozza-Thivent~\cite{cocozza-1985}. It was subsequently studied in detail in the finite one-dimensional lattice with periodic boundary conditions by Evans--Waclaw~\cite{evans-waclaw-2014}. In particular, the Sch\"utz--Sandow model is explicitly discussed in Section~3.1.
The latter establishes necessary and sufficient conditions for the steady state of the misanthrope process to be of product form.}

Stochastic duality for the Sch\"utz--Sandow model has been established in~\cite{carinci-giardina-giberti-redig-2013}. Nonequilibrium properties and hydrodynamics for such a process with open boundaries have been studied in \cite{floreani-redig-sau-2022,franceschini-et-al-2023}.
We note that a different generalization of this model with a different interpretation of the $q$ parameter has been studied by Carinci--Giardin\`a--Redig--Sasamoto~\cite{carinci-giardina-redig-sasamoto-2016} for which they have established self-duality.

In this work, we study 
{a special case of the misanthrope process}
on the ring with two positive parameters; $t$, which governs the speed of the transition, and $q$, which governs the asymmetry in the transitions. 
We call this model the \emph{$(q, t)$ asymmetric simple $K$-exclusion process} or $(q, t)$~$K$-ASEP for short.
The speed will enter through the \emph{$t$-deformed integer} $[n]_t$; see \cref{sec:model}. 
{Using results for the misanthrope process, it will follow that}
the steady state has an exact product form and does not depend on $q$. We will show that the steady state has a nice formula involving $t$-deformed binomial coefficients, also known as \emph{Gaussian polynomials}. 
Further, we will calculate the partition function and demonstrate nontrivial symmetry properties of the steady state weights. We will also study the one-point correlation in detail for different parameters of $t$.

{
As far as we could see, there does not seem to exist a two-dimensional analogue of the misanthrope process on a finite lattice with periodic boundary conditions in the literature.
More precisely, we mean a two-dimensional model where there is an asymmetric rate in one direction and the steady state is of product form that is independent of the asymmetry parameter. In this work, we construct such a two-dimensional process, and we hope this model will prove fruitful in understanding two-dimensional analogues of misanthrope processes.
}

We include simulations for the $(q, t)$~$K$-ASEP with $L = 50$ sites, at most $K = 6$ particles per site and $n = 150$ particles for different values of $q$ and $t$ as ancillary files along with this article. There are nine simulations for $(t, q) \in \{50, 1, 0.02\} \times \{0, 0.5, 1\}$.

We begin with the precise definition of the $(q, t)$~$K$-ASEP and statements of our main results in \cref{sec:model}.

\section{Model description and {summary of results}}
\label{sec:model}

The $(q, t)$ asymmetric simple $K$-exclusion process ($(q, t)$~$K$-ASEP) is defined 
on a periodic one-dimensional lattice by three positive integer parameters -- the number of sites $L$ (sites $1$ and $L$ are adjacent), the maximum number of particles per site $K$, and the total number of particles $n$ satisfying $0 \leq n \leq LK$ -- and two rate parameters $q \geq 0$ and $t > 0$. 
All particles are indistinguishable. We will fix the number of particles per site to be $K$ unless specified otherwise.
We denote the set of configurations by
\begin{equation}
\Omega_{L,n} \equiv \Omega^K_{L,n} = \left\{ \eta \in \{0,1,\dots,K\}^L \biggm| \ds \sum_{i=1}^L \eta_i = n \right\}.
\end{equation}
The number of configurations is thus given by
\begin{equation}
|\Omega_{L,n}| = [x^n] (1 + x + x^2 + \cdots + x^K)^L,
\end{equation}
where $[x^n] f(x)$ denotes the coefficient of $x^n$ in $f(x)$.
For example,
\begin{equation}
\label{conf eg}
\Omega^2_{3,4} = \{ (0, 2, 2), (1, 1, 2), (1, 2, 1), (2, 0, 2), (2, 1, 1), (2, 2, 0) \}
\end{equation}
and $|\Omega^2_{3,4}| = [x^4](1 + x + x^2)^3 = 6$.

\begin{figure}[h!]
\begin{center}

\begin{tikzpicture}
\def\r{3.4}
\def\sr{0.07*\r}
\draw (0,0) circle (\r cm);

\draw[fill=black] (0,\r+\sr) circle (\sr cm);
\draw[fill=black] (0,\r+3*\sr) circle (\sr cm);
\footnotesize
\node(n1) at (0,\r-\sr) {$1$};

\draw[fill=black] ({(\r+\sr)*cos(45)},{(\r+\sr)*sin(45)}) circle (\sr cm);

\node(n1) at ({(\r-1*\sr)*cos(45)},{(\r-\sr)*sin(45)}) {$2$};

\node (G) at ({(\r+3*\sr)*cos(87)},{(\r+3*\sr)*sin(87)}) {};
\node (R) at ({(\r+3*\sr)*cos(48)},{(\r+3*\sr)*sin(48)}) {};
\draw[-{Stealth[length=3mm, width=2mm]}] (G) edge[teal,bend left=20] (R);

\node(n1) at ({(\r+8*\sr)*cos(66)},{(\r+5*\sr)*sin(66)}) {\color{teal}$\qty[2] (\qty[3] - \qty[1])$};

\draw[fill=black] (\r+\sr,0) circle (\sr cm);
\draw[fill=black] (\r+3*\sr,0) circle (\sr cm);
\draw[fill=black] (\r+5*\sr,0) circle (\sr cm);
\node(n1) at ({(\r-1*\sr)*cos(0)},{(\r-\sr)*sin(0)}) {$3$};

\draw[fill=black] ({(\r+\sr)*cos(45)},{-(\r+\sr)*sin(45)}) circle (\sr cm);
\node(n1) at ({(\r-1*\sr)*cos(-45)},{(\r-\sr)*sin(-45)}) {$4$};
\node(n1) at ({(\r-1*\sr)*cos(-90)},{(\r-\sr)*sin(-90)}) {$5$};

\draw[fill=black] ({-(\r+\sr)*cos(45)},{-(\r+\sr)*sin(45)}) circle (\sr cm);
\draw[fill=black] ({-(\r+3*\sr)*cos(45)},{-(\r+3*\sr)*sin(45)}) circle (\sr cm);
\textbf{\node(n1) at ({(\r-1*\sr)*cos(-135)},{(\r-\sr)*sin(-135)}) {$6$};}

\draw[fill=black] (-\r-\sr,0) circle (\sr cm);
\draw[fill=black] (-\r-3*\sr,0) circle (\sr cm);
\node(n1) at ({(\r-1*\sr)*cos(180)},{(\r-\sr)*sin(180)}) {$7$};

\node(n1) at ({(\r-1*\sr)*cos(135)},{(\r-\sr)*sin(135)}) {$8$};

\node (G2) at ({(\r+3*\sr)*cos(93)},{(\r+3*\sr)*sin(93)}) {};
\node (R2) at ({(\r+\sr)*cos(132)},{(\r+\sr)*sin(132)}) {};
\draw[-{Stealth[length=3mm, width=2mm]}] (G2) edge[red,bend right=20] (R2);

\node(n1) at ({-(\r+8*\sr)*cos(66)},{(\r+5*\sr)*sin(66)}) {\color{red}$ q \, \qty[2] (\qty[3] - \qty[0])$};

\end{tikzpicture}
\caption{An example of the configuration $(2, 1, 3, 1, 0, 2, 2, 0) \in \Omega^3_{8, 11}$ and two possible transitions from site $1$.}
\label{fig:eg}
\end{center}
\end{figure}
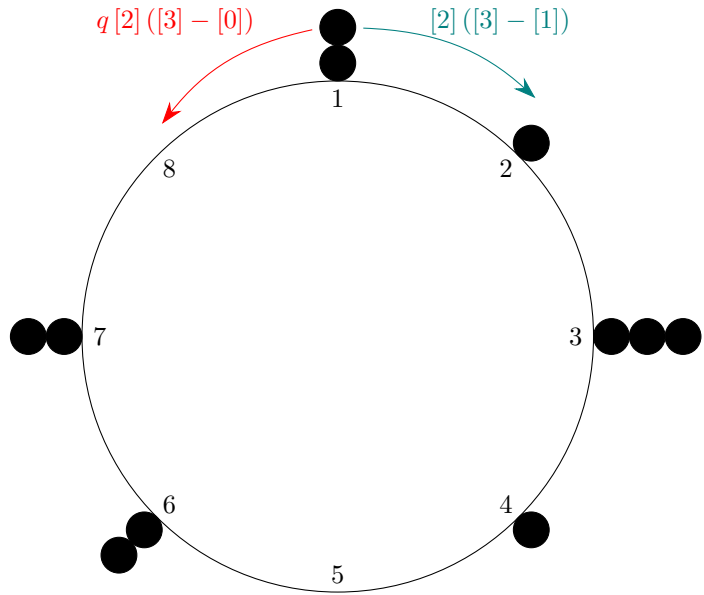

We define the $t$-analog of a nonnegative integer $k$ as 
\begin{equation}
[k] \equiv [k]_t = \sum_{i = 0}^{k - 1} t^i = 
\begin{cases}
\ds \frac{1-t^k}{1-t} & \text{for } t \neq 1,\\
k & \text{for } t=1.
\end{cases}
\end{equation}
We will drop the subscript $t$ for notational convenience.
The transitions are as follows. At each time, one of the particles at a site can either move forward to the site immediately in front of it (i.e. move clockwise) or to the site immediately behind it (i.e. move counterclockwise) with a rate that depends on the number of particles at the origin and target sites. To be precise, consider the configuration $\eta$ and 
focus on the pair of sites $(i, i+1)$ and suppose the number of particles in these sites are $(a, b)$. Then the forward transition is
\begin{equation}
\label{1d-forward}
(a, b) \rightarrow (a-1, b +1 ) \quad \text{with rate} \quad [a] ([K] - [b]),
\end{equation}
and the reverse transition is
\begin{equation}
\label{1d-reverse}
(a, b) \rightarrow (a + 1, b -1 ) \quad \text{with rate} \quad q \, [b] ([K] - [a]).
\end{equation}
Notice that both the transitions are consistent, i.e. they give rate $0$ when the transitions are forbidden. 
When $q = 0$ (resp. $q = 1$), the process becomes totally asymmetric (resp. symmetric).
See \cref{fig:eg} for examples of allowed transitions. When $K = 1$, this corresponds exactly to the single species asymmetric simple exclusion process (ASEP) on the ring, which has been extensively studied. For $K > 1$, this is no longer an exclusion process because there are multiple particles per site. However, this is a simple process because particles jump only to adjacent sites.

This model at $t = q = 1$ has been first studied by Sch\"utz--Sandow~\cite[Section~II.B.]{schutz-sandow-1994}. In that case, the model is reversible.
{As mentioned above, this model at $t = 1$ is discussed in \cite[Section 3.1]{evans-waclaw-2014}. }
Carinci--Giardin\`a--Redig--Sasamoto~\cite[Section 3]{carinci-giardina-redig-sasamoto-2016} studied a different nonreversible generalisation than ours of this model with an extra parameter (which they call $q$). However, their model does not match the $(q, t)$~$K$-ASEP at any value of $q$ and $t$ other than $q = t = 1$.
In both these models, forward and reverse transitions occur. The $(q, t)$~$K$-ASEP has a new parameter which decouples the forward and reverse transitions, and is therefore nonreversible.

It is easy to see that the process is ergodic for $t > 0$ and $q \geq 0$, i.e. there is a sequence of transitions leading from any configuration in $\Omega_{L,n}$ to any other. Therefore, the steady state, which we denote by $\pi$, is unique.
Moreover, the transition rates do not depend on the site positions. Therefore, the process is invariant under cyclic translations.
By ergodicity, the probability of seeing any configuration $\eta \in \Omega_{L,n}$ in the long time limit approaches $\pi(\eta)$. Because of the above two properties, $\pi(\eta) > 0$ for all $\eta$ and  
\begin{equation}
    \pi(\eta_1,\ldots,\eta_L) = \pi(\eta_2,\ldots,\eta_L,\eta_1).
\end{equation}

The $t$-analogue of the factorial is given by
\begin{equation}
   [k]! \equiv [k]_t ! = [1] [2] \cdots [k].
\end{equation}
The \emph{$t$-binomial coefficient} or \emph{Gaussian polynomial} is
\begin{equation}
\qbinom ki \equiv \qbinom ki_t = \frac{[k] !}{[k-i] ! [i] !}.
\end{equation}
For later use, we note the analogue of the Pascal recurrence for the $t$-binomial coefficient,
\begin{equation}
\label{pascal-identity}
\qbinom ki = t^i \qbinom {k-1}i + \qbinom {k-1}{i-1}.
\end{equation}
We will show in \cref{sec:ss} that the \emph{steady state weight} is given by the formula,
\begin{equation}
\label{wt-eta}
\wt_\eta(t) = \prod_{i=1}^L t^{(\eta_i - 1)(\eta_i - 2)/2} \qbinom K{\eta_i},
\end{equation}
so that the \emph{nonequilibrium partition function} or \emph{normalization factor} is
\begin{equation}
\label{pf sum}
Z_{L,n}(t) = \sum_{\eta \in \Omega_{L,n}} \wt_\eta(t).
\end{equation}
Then the steady state probability of $\eta$ is 
\begin{equation}    
\label{steady-state}
\pi(\eta) = \frac{\wt_\eta(t)}{Z_{L,n}}.
\end{equation}
We will give explicit formulas for the grand-canonical partition function in \cref{sec:pf}. Although there is a general formula for the latter in misanthrope processes, we will show that in this case, it has a nice factorized form.

We next study properties of the steady state.
The particle system is clearly translation-invariant since the dynamics in \eqref{1d-forward} and \eqref{1d-reverse} depends only on the number of particles at a given site and not its label. Therefore, the steady state is also translation-invariant. This can be seen for the example in \eqref{conf eg} whose steady state is constructed late in \eqref{ss eg}.

When $t =1$, there is also particle-hole symmetry in the sense that the particle system for parameters $(K, L, n)$ is isomorphic to that for $(K, L, KL - n)$. To see this, note that the forward transition rate becomes
\begin{equation}
\label{1d-forward t=1}
(a, b) \rightarrow (a-1, b +1 ) \quad \text{with rate} \quad a (K - b),
\end{equation}
and thus,
\begin{equation}
(K-b, K-a) \rightarrow (K-(b+1), K-(a-1)) \quad \text{with rate} \quad (K - b) a,
\end{equation}
and similarly for the backward transition. Hence both rates are identical. The isomorphism is therefore given by
taking the configuration $(\eta_1, \dots, \eta_L) \in \Omega_{L,n}$ to  
$(K - \eta_L, \dots, K - \eta_1) \in \Omega_{L, KL - n}$.
For generic values of $t \neq 1$, there is no particle-hole symmetry in the dynamics. 
This can be seen by looking at the eigenvalues of the generators in these two systems, for example.
However, we will show in \cref{sec:ss p-h} that the steady state of the $(q, t)$~$K$-ASEP obeys particle-hole symmetry.

We now want to prove a result about the palindromicity of the steady state weights. Usually, a polynomial $p(x) = \sum_{i = 0}^d c_i x^i$ is said to be palindromic if $c_j = c_{d-j}$ for $0 \leq j \leq d$, or in other words, the coefficients read the same when read left-to-right and right-to-left. We will need a slightly more general definition. Define the \emph{degree} (resp. \emph{order}) of a polynomial $p$ to be the largest (resp. smallest) exponent with nonzero coefficient, denoted $\deg(p)$ (resp. $\ord(p)$). Then, we say that a polynomial $p$ with $\deg(p) = d$ and $\ord(p) = o$, written as $p(x) = \sum_{i = o}^d c_i x^i$, is \emph{palindromic} if $c_j = c_{o + d -j}$ for $o \leq j \leq d$. The advantage of this formulation is that the polynomial $x + 2x^2 + x^3$ is palindromic according to our definition, but not as per the standard one.
It is easy to see that a polynomial with $\deg(p) = d$ and $\ord(p) = o$ is palindromic if and only if
\begin{equation}
\label{palindr}
p\left( \frac{1}{x} \right) = \frac{1}{x^{o + d}} p(x).
\end{equation}
For a palindromic polynomial $p$, we define the \emph{center of mass} to be $(\ord(p) + \deg(p))/2$.

We will prove in \cref{sec:ss pal} that for every $\eta$, $\wt_\eta(t)$ is palindromic in $t$, and moreover that the center of mass of these polynomials is the same, i.e. it depends only on $L, K$ and $n$.
It will thus follow that the sum of steady state weights, which is the partition function $Z_{L,n}(t)$ in \eqref{pf sum}, must certainly be palindromic. 

We now focus on properties of the density for different values of $t$ in \cref{sec:site occ}. 
We will use the notation $\pi(\eta_i = a)$ to denote the probability of having $a$ particles at site $i$ and $\langle \cdot \rangle$ to denote expectations in steady state.
From the formula for the steady state in \eqref{steady-state}, we get that the density is independent of the site and
\begin{equation}
\label{corr-1pt}
\pi(\eta_1 = a) = t^{\frac{(a-1)(a-2)}{2}} \qbinom K{a}  \frac{Z_{L-1,n-a}}{Z_{L,n}}.
\end{equation}
First of all, it is easy to see that the expected value of the density in steady state is
\begin{equation}
\label{exp dens}
\langle \eta_1 \rangle = \frac{n}{L},
\end{equation}
and we sketch the argument at the beginning of \cref{sec:site occ}.
We want to understand the most probable occupation at any site.
This is complicated in general, but certain cases can be analysed. 
We study the case $t = 1$ in \cref{sec:t 1} and show that the most probable density is
\begin{equation}
\label{final ans t=1}
\left\lfloor\frac{(K+1)(n+1)}{KL+2}\right\rfloor,
\end{equation}
which looks quite different from \eqref{exp dens}. 

We next study the most probable density for very large values and very small values of $t$. 
Let $\fracpart{x}$ stand for the fractional part of a real number $x$.
It turns out that when $\fracpart{n/L} = 1/2$, the analysis becomes much more complicated. So, for these two cases, we assume that $\fracpart{n/L} \neq 1/2$. We will show a rather surprising result about these two cases in \cref{sec:t large} and \cref{sec:t small}. By very different computations that the most probably density occurs at exactly the same value, namely
\begin{equation}
\label{extreme_val_ast}
a^{\ast}=\begin{cases}
	\left\lfloor\dfrac{n}{L}\right\rfloor, 
	& \text{if $\ds\fracpart{\dfrac{n}{L}} < \frac 12 $}, \\   \noalign{\vskip9pt}
    \left\lceil\dfrac{n}{L}\right\rceil, 
    & \text{if $\ds\fracpart{\dfrac{n}{L}} > \frac 12 $},
\end{cases}
\end{equation}
with probability of occupation 
\begin{equation}
\label{extreme_val_prob}
\pi(\eta_i = a^\ast) = 
\begin{cases}
\ds 1 - \fracpart{\frac{n}{L}} & \text{if  $\ds \fracpart{\dfrac{n}{L}} \leq \frac{1}{2}$},\\
\noalign{\vskip9pt}
\ds \fracpart{\frac{n}{L}} & \text{if  $\ds \fracpart{\dfrac{n}{L}} > \frac{1}{2}$},
\end{cases}
\end{equation}
which is always at least $1/2$. 

We compare these results for $t = 1$, very large, and very small values of $t$ when $\fracpart{n/L} \neq 1/2$ in \cref{sec:t comp}. We show that the most probable density for $t = 1$ in \eqref{final ans t=1}, and that for large and small values of $t$ in \eqref{extreme_val_ast} differ by at most $1$, and thus the most probable density can switch at most two times as we increase $t$ from very small to very large values.

We end the discussion of the density by some conjectures about the most probable density at $\fracpart{n/L} = 1/2$ for very small and very large values of $t$ in \cref{sec:n/L=1/2}, and give some evidence in \cref{sec:eg}.

Finally, in \cref{sec:2d}, we study a two-dimensional exclusion process on a discrete cylinder with circumference $L$ and height $K$, and having $n$ particles. 
We believe this process is interesting in its own right.
We will show that the steady state of this process is of product form in \cref{sec:2dss}, and that it projects (as a stochastic process) to the $(q, t)$~$K$-ASEP in \cref{sec:proj}.
{As a consequence, the steady state of the $(q, t)$~$K$-ASEP can be obtained as a sum over steady state weights of the two-dimensional exclusion process, and we illustrate that as well.}

\section{Steady state distribution}
\label{sec:ss}

{To prove the formula \eqref{steady-state} for the steady state, we first verify that this is a misanthrope process using \cite[Equations (14) and (15)]{evans-waclaw-2014}. If $u(m, n)$ is the rate at which a particle in a site containing $m$ particles moves to the neighbouring site containing $n$ particles, then the conditions are given by
\begin{equation}
\begin{split}
\label{misanthrope}
\frac{u(n, m)}{u(m+1, n-1) } &= \frac{u(1, m) u(n, 0)}
{u(m+1, 0) u(1, n-1)}, \\
u(n, m) - u(m, n) &= u(n, 0) - u(m, 0).
\end{split}
\end{equation}
We first look at the case $q = 0$. In that case $u(m, n) = [m]([K] - [n])$ and it is an easy exercise to verify that \eqref{misanthrope} holds for this choice of rates. Therefore, the $(q, t)$~$K$-ASEP is indeed a special case of the misanthrope process and therefore the steady state is of product form.}

{We now verify the formula \eqref{steady-state} when $q = 0$. That is, we have to show that the probability of having $a$ particles at any site in the steady state is
\[
f(a) = t^{(a - 1)(a - 2)/2} \qbinom K{a}.
\]
Let $y_i = u(i, 0) = [i][K]$ and $x_i = u(1, i) = [K] - [i] = t^i [K-i]$. Then, using~\cite[Equation (22)]{evans-waclaw-2014}, 
\[
f(a) = \prod_{i = 1}^a \frac{x_{i-1}}{y_i},
\]
we obtain, after plugging in and simplifying,
\begin{equation}
\label{density}
f(a) = \frac{t^{a(a - 1)/2}}{[K]^a} \qbinom K{a}.
\end{equation}
This is equivalent to the above formula because their ratio is $t^{-a + 1}/[K]^a$, and since the total number of particles is conserved, this factor is constant for all configurations, and can be discarded. 
}

{It is a standard argument for misanthrope processes to show that if the steady state is of product form, then it will be independent of the asymmetry parameter. We repeat it below for completeness.
We can write the generator $M$ of the $(q, t)$~$K$-ASEP as the column-stochastic matrix $M = M_0 + q M_1$, where $M_0$ is the generator corresponding to $q = 0$ and $M_1$ encodes the reverse transitions. Notice that $M_0$ and $M_1$ are of the same form. Let $v$ be the column vector encoding the steady state. Then we have shown that $M_0 \cdot v = 0$. Since $v$ is of product form, the same argument as above will show that $M_1 \cdot v = 0$, and therefore, $M \cdot v = 0$. Thus, $v$ also encodes the steady state for the $(q, t)$~$K$-ASEP.
}

For the example with $K = 2$, $L = 3$ and $n = 4$ considered in \eqref{conf eg}, the steady state distribution is given by
\begin{multline}
\label{ss eg}
\pi(0, 2, 2) = \frac{t}{Z^2_{3,4}}, 
\pi(1, 1, 2) = \frac{(t+1)^2}{Z^2_{3,4}}, 
\pi(1, 2, 1) = \frac{(t+1)^2}{Z^2_{3,4}}, \\
\pi(2, 0, 2) = \frac{t}{Z^2_{3,4}}, 
\pi(2, 1, 1) = \frac{(t+1)^2}{Z^2_{3,4}}, 
\pi(2, 2, 0) = \frac{t}{Z^2_{3,4}},
\end{multline}
where
\begin{equation}
\label{pf eg}
Z^2_{3,4} = 3 \left(t^2+3 t+1\right).
\end{equation}

\subsection{Partition function}
\label{sec:pf}

Define the \emph{grand-canonical partition function}
\begin{equation}
\cZ_L(y) = \sum_{n = 0}^{K L} Z_{L,n} y^n.
\end{equation}
{As shown in \cite[Equation (30)]{evans-waclaw-2014}, $\cZ_L(y)$ can be written as the $L$'th power of $\sum_{m \geq 0} f(m)y^m$. In our case, this is a finite sum, and we obtain, using \eqref{density},}
\begin{equation}
\label{cZ-frac1}
\cZ_L(y) = \left( \sum_{i=0}^K  y^i t^{\frac{(i-1)(i-2)}{2}} \qbinom Ki \right)^L.
\end{equation}
It turns out that the sum can be simplified further. A standard result in combinatorics is the $t$-binomial theorem~\cite[Equation (1.87)]{stanley-ec1}, which says
\begin{equation}
\label{q-binomial}
\sum_{j=0}^m t^{j(j-1)/2} \qbinom mj_t z^j = \prod_{j=0}^{m-1} (1 + t^j z).
\end{equation}
Setting $t=1$ gives the binomial theorem and hence this is a natural generalization.
We now claim that the sum in \eqref{cZ-frac1} can be written as
\begin{equation}
\label{pf-identity}
\sum_{i=0}^K  y^i t^{\frac{(i-1)(i-2)}{2}} \qbinom Ki
= t \prod_{i = -1}^{K-2} (1 + t^i y)
= (y + t) \prod_{i=0}^{K-2} (1 + t^i y),
\end{equation}
for $K \geq 1$. We will prove this by induction on $K$. When $K = 1$, the product on the right hand side is $1$ vacuously and so the result holds trivially. 
Now suppose $K \geq 2$. Expanding the right hand side of \eqref{pf-identity} using \eqref{q-binomial}, we have
\begin{multline}
(y + t) \prod_{i=0}^{K-2} (1 + t^i y) 
= (y + t) \sum_{i = 0}^{K-1} t^{i(i-1)/2} y^i \qbinom {K-1}i \\
= \sum_{i = 0}^{K-1} t^{i(i-1)/2 + 1} y^i \qbinom {K-1}i
+ \sum_{i = 1}^{K} t^{(i-1)(i-2)/2} y^i \qbinom {K-1}{i-1} \\
= \sum_{i = 0}^{K} t^{(i-1)(i-2)/2} y^i \left( t^i \qbinom {K-1}i + \qbinom {K-1}{i-1} \right).
\end{multline}
Now use \eqref{pascal-identity} to get the left hand side of \eqref{pf-identity}, completing the proof. Substituting \eqref{pf-identity} in the formula \eqref{cZ-frac1} for the generating function for the partition function, we finally obtain the compact formula
\begin{equation}
\label{cZ-frac2}
\cZ_L(y) = t^L \prod_{i = -1}^{K-2}(1 + t^i y)^L.
\end{equation}
The same proof shows that if we assign \emph{activities} $x_j$ for the number of particles at site $j$ so that the 
probability of having $i$ particles at site $j$ is proportional to $x_j^i t^{\frac{(i-1)(i-2)}{2}} \qbinom Ki$, then the modified grand-canonical partition function defined by
\begin{equation}
\cZ_L(x_1, \dots, x_L) = \sum_{n = 0}^{K L} \sum_{\eta \in \Omega_{L,n}} 
\prod_{j = 1}^L
x_j^{\eta_j} t^{\frac{(\eta_j-1)(\eta_j-2)}{2}} \qbinom K {\eta_j}
\end{equation}
simply becomes
\begin{equation}
\label{pf-gen fn}
\cZ_L(x_1, \dots, x_L) = t^L \prod_{j = 1}^L \prod_{i = -1}^{K-2}
(1 + t^i x_j).
\end{equation}

In the special case $t = 1$, the formula for the partition function $Z_{L,n}$ simplifies considerably. The grand-canonical partition function in \eqref{cZ-frac2} becomes
\begin{equation}
\cZ_L(y) = (1 + y)^{K L},
\end{equation}
and by the binomial theorem, that shows that
\begin{equation}
Z_{L,n} = \binom{KL}{n},
\end{equation}
which is also written down in \cite[Equation (2.31)]{schutz-sandow-1994}.

\subsection{Particle-hole symmetry in the steady state}
\label{sec:ss p-h}

 We want to show that $(\eta_1, \dots, \eta_L)$ and $(K - \eta_1, \dots, K - \eta_L)$ have the same probability in steady state.
To see this, recall from \eqref{steady-state} that
\begin{equation}
\pi(K - \eta_1, \dots, K - \eta_L) \propto 
\prod_{i=1}^L t^{(K - \eta_i - 1)(K - \eta_i - 2)/2} \qbinom K{K - \eta_i}.
\end{equation}
But since $\qbinom K{K - \eta_i} = \qbinom K{\eta_i}$, we only need to compare the powers of $t$. A short computation shows that
\begin{multline}
\sum_{i=1}^L \left( \frac{(\eta_i - 1)(\eta_i - 2)}2 - 
\frac{(K - \eta_i - 1)(K - \eta_i - 2)}2 \right) \\
= (K-3) \left(n - \frac{KL}{2} \right), 
\end{multline}
which does not depend on the configuration $\eta$.
Thus, the steady state weights of both configurations are proportional to the same power of $t$ and this will get washed away in the steady state probabilities.

To compare with \eqref{ss eg}, consider the system with $K = 2, L = 3$ and $n = 2$. The steady state is then
\begin{multline}
\label{ss eg2}
\pi(0, 0, 2) = \frac{t^2}{Z^2_{3,2}}, 
\pi(0, 1, 1) = \frac{t(t+1)^2}{Z^2_{3,2}}, 
\pi(0, 2, 0) = \frac{t^2}{Z^2_{3,2}}, \\
\pi(1, 0, 1) = \frac{t(t+1)^2}{Z^2_{3,2}}, 
\pi(1, 1, 0) = \frac{t(t+1)^2}{Z^2_{3,2}}, 
\pi(2, 0, 0) = \frac{t^2}{Z^2_{3,4}},
\end{multline}
and
\begin{equation}
Z^2_{3,2} = t \, Z^2_{3,4} = 3t \left(t^2+3 t+1\right).
\end{equation}

\subsection{Palindromicity}
\label{sec:ss pal}

To prove the palindromicity of $\wt(\eta)$, we first consider the $t$-binomial coefficient. From the definition
\begin{equation}
\label{t-binom}
\qbinom K{a} = \frac{(1+t+\cdots+t^{K-1}) (1+t+\cdots+t^{K-2}) \dots (1 + t + \cdots + t^a)} {(1+t+\cdots+t^{K-a-1}) \dots (1+t)(1)}.
\end{equation}
Therefore, the highest power of $t$  is
\begin{equation}
\label{deg-tbinom}
\deg \left(\qbinom K{a} \right) = \frac{K(K-1)}{2} - \frac{(K-a)(K-a-1)}{2} - \frac{a(a-1)}{2}
\end{equation}
and the lowest power of $t$ is
\begin{equation}
\label{ord-tbinom}
\ord \left(\qbinom K{a} \right) = 0.
\end{equation}
It is an easy exercise to verify, using \eqref{t-binom} for example, that
\begin{equation}
\qbinom K{a}_{1/t} = t^{\frac{(K-a)(K-a-1)}{2} + \frac{a(a-1)}{2} - \frac{K(K-1)}{2}} \qbinom K{a}_t,
\end{equation}
and therefore the $t$-binomial coefficient is palindromic by \eqref{palindr}.

Now, we look at the weights in \eqref{wt-eta}. First, the order is
\begin{equation}
\label{ord-wt}
\ord(\wt_\eta) = \sum_{i = 1}^L \frac{(\eta_i - 1)(\eta_i - 2)}{2}
= \frac{1}{2} \sum_{i = 1}^L \eta_i^2 - \frac{3}{2}n + L.
\end{equation}
Secondly, the degree is 
\begin{multline}
\label{deg-wt}
\deg(\wt_\eta) = \\
\ord(\wt_\eta) + \sum_{i = 1}^L \left( \frac{K(K - 1)}{2} 
- \frac{\eta_i(\eta_i - 1)}{2} 
- \frac{(K - \eta_i)(K - \eta_i - 1)}{2} \right) \\
= -\frac{1}{2} \sum_{i = 1}^L \eta_i^2 - \frac{3}{2}n + Kn + L,
\end{multline}
and thus, their center of mass is
\begin{equation}
\label{ord+deg-wt}
\frac{\ord(\wt_\eta) + \deg(\wt_\eta)}2 = L + \frac{(K - 3)n}2.
\end{equation}
Now look at 
\begin{equation}
\wt_\eta \left( \frac{1}{t} \right) 
= \prod_{i=1}^L t^{-(\eta_i - 1)(\eta_i - 2)/2
-K(K - 1)/2 + \eta_i(\eta_i - 1)/2 
+ (K - \eta_i)(K - \eta_i - 1)/2} \qbinom K{\eta_i}_t,
\end{equation}
which, after simplification becomes
\begin{equation}
\wt_\eta \left( \frac{1}{t} \right) 
= t^{\sum_{i=1}^L \left((3 - K)\eta_i -2 \right)} \wt_\eta(t)
= t^{(3-K)n - 2L} \wt_\eta(t).
\end{equation}
Comparing with \eqref{ord+deg-wt} and using the property in \eqref{palindr}, we see that $\wt(\eta)$ is palindromic for all $\eta$. This can be seen in the example in \eqref{ss eg}.

In general, the sum of palindromic polynomials is not palindromic. For example, the sum of $1 + t$ and $t + t^2 + t^3$ is not palindromic. However, if palindromic polynomials have the same center of mass, their sum is clearly palindromic.
We have shown in \eqref{ord+deg-wt} that all steady state weights have the same center of mass and therefore any linear combination of these weights is palindromic. Therefore, any probability in steady state has numerator equal to a palindromic polynomial. Moreover, the expectation of any quantity which itself is not a function of $t$ has the same property.

In particular, the sum of steady state weights, which is the partition function $Z_{L,n}(t)$ in \eqref{pf sum}, must certainly be palindromic. 
The degree and order of the partition function are computed later in \eqref{deg-Z} and \eqref{ord-Z} respectively, and one can check that its center of mass is also equal to \eqref{ord+deg-wt}.
Again, this is demonstrated in \eqref{pf eg}. Note that a direct proof of this latter statement can be given using \eqref{cZ-frac2}.

By palindromicity of the steady state weights, it is clear that the steady state is invariant if we replace $t$ by $1/t$. However, the dynamics itself is not invariant under this transformation. In other words, the Markovian dynamics of the system with parameter $t$ is not isomorphic to that with parameter $1/t$. This can again be verified by looking at the eigenvalues of the generators in these two systems, for example.

\section{Probability of occupation numbers at a site}
\label{sec:site occ}

The expected number of particles in any site is
\begin{equation}
\langle \eta_1 \rangle = \sum_{a = 0}^K a\, \pi(\eta_1 = a).
\end{equation}
By translation invariance of the dynamics mentioned in the beginning of \cref{sec:ss p-h}, $\langle \eta_1 \rangle = \langle \eta_i \rangle$ for all $i$ and since there are a total of $n$ particles, we get that
\begin{equation}
\langle \eta_1 \rangle = \frac{n}{L}.
\end{equation}

\subsection{$t = 1$}
\label{sec:t 1}

In this case, we already know from \eqref{corr-1pt} that
\begin{equation}
\label{t1_most probable}
\pi(\eta_1 =a) = \binom{K}{a} \frac{\ds\binom{K(L-1)}{n-a}}{\ds\binom{KL}{n}}.
\end{equation}
The ratio 
\begin{equation}
\frac{\pi(\eta_1 =a)}{\pi(\eta_1 =a+1)} = \frac{(a+1)(KL-K-n+a+1)}{(K-a)(n-a)},
\end{equation}
from which we get that the most probable value occurs at
\begin{equation}
a^{\ast} = \left\lceil\frac{(K+1)(n+1)}{KL+2} -1 \right\rceil = \left\lfloor\frac{(K+1)(n+1)}{KL+2}\right\rfloor.
\end{equation}
{
We can show (see \eqref{difference atmost 1 a} and \eqref{difference atmost 1 b} in \cref{sec:t comp}) that $a^\ast$ can only take two values given by 
\begin{equation}
a^{\ast}=\begin{cases}
		\left\lfloor\dfrac{n}{L}\right\rfloor, & \text{if  $\ds (K+1)\fracpart{\dfrac{n}{L}}+(L-2)\left\lfloor\dfrac{n}{L}\right\rfloor < KL-K+1$},\\
       \noalign{\vskip9pt}
                \left\lceil\dfrac{n}{L}\right\rceil, & \text{if  $\ds (K+1)\fracpart{\dfrac{n}{L}} + (L-2)\left\lfloor\dfrac{n}{L}\right\rfloor \geq KL-K+1$},
    		 \end{cases}
\end{equation}
Plugging this value into \eqref{t1_most probable} gives the most probable occupation probability.
}

\subsection{Large $t$ when $\fracpart{n/L} \neq 1/2$}
\label{sec:t large}

Here we will assume that $t$ is large enough so that only the highest power of $t$ contributes to any polynomial in $t$. 
Let $a^{\ast}$ be the most probably occupation number for a given system with system size $L$, maximum number of particles per site $K$ and total number of particles $n$. There are four factors in the numerator of \eqref{corr-1pt} which depend on $t$. The first is the power of $t$, which we set aside for now. The second is the $t$-binomial coefficient given in \eqref{t-binom}, whose degree is given in \eqref{deg-tbinom}.
Now to find the highest power of $t$ in $Z_{L-1,n-a}$ we use \eqref{cZ-frac2}.
Since we want to extract the coefficient of $y^n$ while getting the maximum power of $t$, we take 
$k = \lfloor n/L \rfloor$ terms of the form $(1 + t^i y)^L$ for the largest values of $i$ and the remaining from the next largest.
This gives the degree of $Z_{L,n}(t)$ as
\begin{equation}
\label{deg-Z}
    \begin{split}
\deg(Z_{L,n})  &= L + \frac{kL}{2}((K-2)+(K-(k+1))) + (K-(k+2))(n-kL)\\
    &= (k^2+k+2)\frac{L}{2} + (K-k-2)n.
    \end{split}
\end{equation}
Therefore, the degree of $Z_{L-1,n-a}(t)$ is 
\begin{equation}
\deg(Z_{L-1,n-a})  = (k_2(a)^2+k_2(a)+2)\frac{L-1}{2} + (K-k_2(a)-2)(n-a),
\end{equation}
where $k_2(a) = \left\lfloor (n-a)/(L-1)\right\rfloor$.

Combining these pieces, we see that the highest power of $t$ in $\pi(\eta_i=a)$ is
\begin{multline}
h(a) = \frac{(a-1)(a-2)}{2} + \frac{K(K-1)}{2} - \frac{(K-a)(K-a-1)}{2} - \frac{a(a-1)}{2} \\
+(k_2(a)^2+k_2(a)+2)\frac{L-1}{2} + (K-k_2(a)-2)(n-a) -(k^2+k+2)\frac{L}{2} - (K-k-2)N,
\end{multline}
which simplifies to 
\begin{multline}
\label{h fn}
h(a) =  \frac{(L-1)}{2} \left\lfloor \dfrac{n-a}{L-1} \right\rfloor
\left( \left\lfloor\dfrac{n-a}{L-1}\right\rfloor+1 \right) - \left\lfloor\dfrac{n-a}{L-1}\right\rfloor (n-a) \\
- \frac{L}{2} \left\lfloor \dfrac{n}{L} \right\rfloor
\left(\left\lfloor \dfrac{n}{L} \right\rfloor+1 \right)
+n\left\lfloor\dfrac{n}{L}\right\rfloor - \frac{a(a-1)}{2}.
\end{multline}
Notice that $h(a)$ is almost a quadratic function of $a$ (except for the floor functions). To find the maximum, we approximate it by a continuous function and elementary calculus shows that the maximum occurs at
$n / L$.
Thus, the maximum occurs at either $\left\lfloor n /L \right\rfloor$  or $\left\lceil n/L \right\rceil$.
Plugging in both these values into \eqref{h fn}, we get that the maximum value is
\begin{equation}
\label{hval_ast}
a^{\ast}=\begin{cases}
	\left\lfloor\dfrac{n}{L}\right\rfloor, 
	& \text{if $\ds\fracpart{\dfrac{n}{L}} < \frac 12 $}, \\   \noalign{\vskip9pt}
    \left\lceil\dfrac{n}{L}\right\rceil, 
    & \text{if $\ds\fracpart{\dfrac{n}{L}} > \frac 12 $}.
\end{cases}
\end{equation}
Extracting the coefficients of the highest power of $t$ from each term, we end up with 
\begin{equation}
\label{effec dens t large}
    \pi(\eta_1=a) = \ds\frac{\ds\binom{L-1}{n-a- k_2(a) (L-1)}}{\ds\binom{L}{n- k L}} t^{h(a)}.
\end{equation}
For the values of $a^\ast$ given in \eqref{hval_ast} we get after a short calculation that $h(a^{\ast}) = 0$, which is expected since otherwise the formula in \eqref{effec dens t large} cannot be a probability for large $t$.
For all other values of $a$, $h(a)$ will be negative. With $t$ being very large, having a negative power of $t$ will make the probability very small and the dominant probability will be $\pi(\eta_1 = a^{\ast})$ given by
\begin{equation}
\label{final ans t large}
\ds\frac{\ds\binom{L-1}{n - a^\ast - k_2(a^\ast) (L-1)}}{\ds\binom{L}{n- k L}}
= \begin{cases}
\ds 1 - \fracpart{\frac{n}{L}} & \text{if  $\ds \fracpart{\dfrac{n}{L}} \leq \frac{1}{2}$},\\
\noalign{\vskip9pt}
\ds \fracpart{\frac{n}{L}} & \text{if  $\ds \fracpart{\dfrac{n}{L}} > \frac{1}{2}$},
\end{cases}
\end{equation}
which is always at least half.

\subsection{Small $t$ when $\fracpart{n/L} \neq 1/2$}
\label{sec:t small}

We repeat the calculation for the most likely value of the density when $t$ is small enough so that only the lowest power of $t$ contributes. 
The ideas are very similar to those in \cref{sec:t large}, and so we will only highlight the main ideas.

The order of the $t$-binomial coefficient in \eqref{t-binom} is given in \eqref{ord-tbinom}.
For the partition function, we now want to collect the coefficient of $y^n$ from \eqref{cZ-frac2} with the lowest powers of $t$. As before, $k = \lfloor n/L \rfloor$ terms will be chosen for the smallest values of $i$ and the remaining from the next smallest. This gives the order of $Z_{L,n}$ as
\begin{equation}
\label{ord-Z}
    \begin{split}
\ord(Z_{L,n}) &= L + \frac{kL}{2}(-1+(k-2)) + (k-1)(n-kL)\\
    &= (-k^2-k+2)\frac{L}{2} + (k-1)n,
    \end{split}
\end{equation}
and the order of $Z_{L-1,n-a}$ as
\begin{equation}
\ord(Z_{L-1,n-a}) = (-k_2(a)^2-k_2(a)+2)\frac{L-1}{2} + (k_2(a)-1)(n-a),
\end{equation}
where, as before, $k_2(a) = \left\lfloor (n-a)/(L-1)\right\rfloor$.
The lowest power of $t$ in $\pi(\eta_1=a)$ is thus
\begin{multline}
\ell(a) = \frac{(a-1)(a-2)}{2} + (-k_2(a)^2-k_2(a)+2)\frac{L-1}{2} + \\
(k_2(a)-1)(n-a) - (-k^2-k+2)\frac{L}{2} - (k-1)n,
\end{multline}
which simplifies to
\begin{multline}
\label{l fn}
\ell(a) = \frac{(a-1)(a-2)}{2} + \frac{1}{2} \left\lfloor\dfrac{n-a}{L-1}\right\rfloor \left( 2n-2a-(L-1) \left(\left\lfloor\dfrac{n-a}{L-1}\right\rfloor+1 \right) \right) \\ 
- \frac{1}{2}\left\lfloor\dfrac{n}{L}\right\rfloor 
\left( 2n-L \left(\left\lfloor\dfrac{n}{L}\right\rfloor+1 \right) \right).
\end{multline}
Again, this is almost a quadratic function of $a$ and we find the maximum as before by approximating it by a continuous function. Elementary calculus shows that the maximum occurs at $n / L$.
Thus, the maximum occurs at either $\left\lfloor n /L \right\rfloor$  or $\left\lceil n/L \right\rceil$.
Plugging in both these values into \eqref{l fn}, we see that the maximum occurs at 
\begin{equation}
\label{lval_ast}
a^{\ast}=\begin{cases}
		\left\lfloor\dfrac{n}{L}\right\rfloor, & \text{if  $\ds \fracpart{\dfrac{n}{L}} < \frac{1}{2}$},\\
       \noalign{\vskip9pt}
                \left\lceil\dfrac{n}{L}\right\rceil, & \text{if  $\ds \fracpart{\dfrac{n}{L}} > \frac{1}{2}$},
    		 \end{cases}
\end{equation}
which is exactly the same as that for $t \gg 1$ computed in \eqref{hval_ast}.
Extracting the coefficients of the lowest power of $t$ from each term we end up with 
\begin{equation}
    \pi(\eta_1=a) = \ds\frac{\ds\binom{L-1}{n-a- k_2(a) (L-1)}}{\ds\binom{L}{n- k L}} t^{\ell(a)}.
\end{equation}
Arguing as before, we get that $\ell(a^{\ast}) = 0$, which can also be verified after a short calculation. All other values of $a$ will have $\ell(a) > 0$ which will ensure that their probabilities are small since $t \ll 1$. Finally, the dominant probability turns out also to be exactly the same as for very large $t$ values given in \eqref{final ans t large}.

\subsection{Comparison of the results for $t \ll 1, t = 1$ and $t \gg 1$}
\label{sec:t comp}

We focus on the case when $\fracpart{n/L} \neq 1/2$. When $t = 1$, we always have that the most probable value of the density is given by 
\eqref{final ans t=1}, whereas for both $t \ll 1$ and $t \gg 1$, this is given by \eqref{final ans t large} and this depends on whether $\fracpart{n/L}$ is more or less than $1/2$. 

\begin{figure}[h!]
\begin{center}
\includegraphics[scale=0.6]{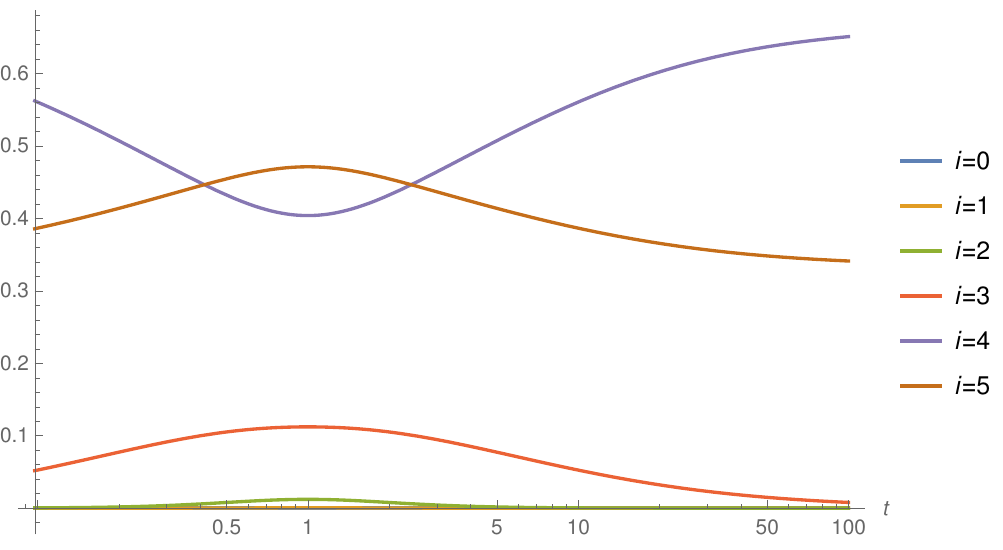}
\end{center}
\caption{For the system with $L = 9, K = 5$ and $n = 39$ particles, the probabilities $\pi(\eta_1 = i)$ of having $i$ particles at the first site as a function of $t$ for $t \in [0, 100]$ and for $0 \leq i \leq 5$. Note that the $x$-axis is drawn on a logarithmic scale. The maximum for $t = 1$ occurs at $i=5$ whereas for $t \ll 1$ and $t \gg 1$, it occurs at $i=4$.
}
\label{fig:densityeg}
\end{figure}

It is natural to ask what happens for intermediate values of $t$.
What we will show now is that the two values differ by at most $1$. First, suppose $\fracpart{n/L} < 1/2$.
Using our previous notation, $n = kL+r$ where $0\leq r < \frac{L}{2}$ and $k \leq K$. The difference is then 
\begin{equation}
\label{difference atmost 1 a}
\begin{split}
    &\left\lfloor\frac{(K+1)(n+1)}{KL+2}\right\rfloor - \left\lfloor\frac{n}{L}\right\rfloor =  \left\lfloor\frac{(K+1)(kL+r+1)}{KL+2}\right\rfloor - k\\
&= \left\lfloor\frac{(K+1)(r+1)+(L-2)k}{KL+2}\right\rfloor.
\end{split}
\end{equation}
This is increasing both as a function of $r$ and $k$. Thus, the minimum value occurs at $r=k=0$ and gives $0$ and the maximum
occurs at $r = L/2$ and $k=K$. For $K \geq 1$ and $L \geq 2$, we have
\begin{equation}
    KL+2 \leq (K+1)\left( \frac{L}{2}+1 \right) + K(L-2) < 2(KL+2)
\end{equation}
and thus
\begin{equation}
    \left\lfloor\frac{(K+1)(L/2 + 1) + K(L-2)}{KL+2}\right\rfloor = 1,
\end{equation}
proving the claim in this case. See \cref{fig:densityeg} for an illustration for when this difference is $1$. 

Now, suppose $\fracpart{n/L} > 1/2$. With $n = kL+r$ again, we have $L/2 < r < L$ and $k < K$. 
\begin{equation}
\label{difference atmost 1 b}
\begin{split}
    \left\lceil\frac{n}{L}\right\rceil - \left\lfloor\frac{(K+1)(n+1)}{KL+2}\right\rfloor =& k+1 - \left\lfloor\frac{(K+1)(kL+r+1)}{KL+2}\right\rfloor\\
    =&1- \left\lfloor\frac{(K+1)(r+1)+(L-2)k}{KL+2}\right\rfloor.
\end{split}
\end{equation}
This is a decreasing function of both $r$ and $k$.
For the maximum value, we take $r = L/2$ and $k=0$. Since, for $K>1$ and $L>1$ we have the inequality $0 \leq (K+1)(L/2 +1) < KL+2$, and
\begin{equation}
    1- \left\lfloor\frac{(K+1)(L/2+1)}{KL+2}\right\rfloor = 1
\end{equation}
Similarly, the minimum value occurs at $r=L-1$ and $k=K-1$. Again, for $K>1$ and $L>1$ we have
\begin{equation}
KL+2 \leq (K+1)(L-1+1) + (L-2)(K-1) < 2(KL+2).
\end{equation}
Thus,
\begin{equation}
    1- \left\lfloor\frac{(K+1)L + (L-2)(K-1)}{KL+2}\right\rfloor = 0,
\end{equation}
completing the proof.

\subsection{At $\fracpart{n/L} = 1/2$ for $t \ll 1$ and $t \gg 1$}
\label{sec:n/L=1/2}

From our previous notation $n = k L+r$, $\fracpart{n/L} = 1/2$ can happen only if $r = L/2$ and since $r$ has to be a non negative integer this tells us that $L$ must be even for this case to happen. We also have the condition $0 \leq k \leq K-1$.
We focus on when $t \ll 1$ and $t \gg 1$. From the discussion in \cref{sec:t large,sec:t small}, we know that the most likely number of particles is either $k$ or $k+1$. The question we want to address is which one is most likely.

For $L = 2$, it is not difficult to see that the numerators of $\pi(\eta_1 = k)$ and $\pi(\eta_1 = k+1)$ are equal. In that case, $n = 2k + 1$, with $0 \leq k \leq K-1$. Then, $Z_{1,k} = t^{(k-1)(k-2)/2} \qbinom Kk$ using \eqref{cZ-frac2} and so, 
\begin{equation}
\pi(\eta_1 = k) \propto t^{(k-1)(k-2)/2} \qbinom Kk Z_{1, k+1}
= t^{k(k-1)/2} \qbinom K{k+1} Z_{1, k} \propto \pi(\eta_1 = k+1). 
\end{equation}

We know from the discussion in \cref{sec:ss p-h} that the numerators of $\pi(\eta_1 = k)$ and $\pi(\eta_1 = k+1)$ are palindromic.
We now claim that, for $L > 2$,
\begin{enumerate}
\item 
\label{it:1}
the top (and bottom) $k+1$ coefficients of these polynomials are equal, 

\item 
\label{it:2}
the $(k+2)$'th coefficient for $\pi(\eta_1 = k)$ is larger (resp. smaller) if $n < K L/2$ (resp. $n > K L/2$), and

\item both are equal polynomials if $n = K L /2$.
\end{enumerate}

The third point is clear from the particle-hole symmetry proved in \cref{sec:ss p-h}. It does not seem to be very easy, nor is it particularly interesting, to prove the other two properties and we illustrate them with a nontrivial example in \cref{sec:eg}.

\section{Two-dimensional enriched process}
\label{sec:2d}

The $(q, t)$~$K$-ASEP can be extended to a two-dimensional exclusion process on a cylinder with circumference $L$ and height $K$, and having $n$ particles, which we now describe. 
We will denote particles by $\occ$ and vacancies by $\vac$.
Since this is an exclusion process, each site will contain at most one particle.
The set of configurations will be denoted
\begin{equation}
\widehat{\Omega}_{L,n} \equiv \widehat{\Omega}^K_{L,n} = \left\{\hat\eta \in \{\vac,\occ\}^{\{1, \dots, K\} \times \{1, \dots, L\}}
\biggm| \ds \sum_{i = 1}^{K} \sum_{j = 1}^L \hat\eta_{i,j} = n \right\},
\end{equation}
where we index the columns from top to bottom as in matrix notation.
For example, $\widehat\Omega^2_{3,4}$ is
\begin{equation}
\label{2d enriched example}
\resizebox{0.95\textwidth}{!}{$
\left\{
\begin{array}{ccccc}
\left(
\begin{array}{ccc}
 \vac & \vac & \occ \\
 \occ & \occ & \occ \\
\end{array}
\right),
&
\left(
\begin{array}{ccc}
 \vac & \occ & \vac \\
 \occ & \occ & \occ \\
\end{array}
\right),
&
\left(
\begin{array}{ccc}
 \vac & \occ & \occ \\
 \vac & \occ & \occ \\
\end{array}
\right),
&
\left(
\begin{array}{ccc}
 \vac & \occ & \occ \\
 \occ & \vac & \occ \\
\end{array}
\right),
&
\left(
\begin{array}{ccc}
 \vac & \occ & \occ \\
 \occ & \occ & \vac \\
\end{array}
\right),
\\[0.3cm]
\left(
\begin{array}{ccc}
 \occ & \vac & \vac \\
 \occ & \occ & \occ \\
\end{array}
\right),
&
\left(
\begin{array}{ccc}
 \occ & \vac & \occ \\
 \vac & \occ & \occ \\
\end{array}
\right),
&
\left(
\begin{array}{ccc}
 \occ & \vac & \occ \\
 \occ & \vac & \occ \\
\end{array}
\right),
&
\left(
\begin{array}{ccc}
 \occ & \vac & \occ \\
 \occ & \occ & \vac \\
\end{array}
\right),
&
\left(
\begin{array}{ccc}
 \occ & \occ & \vac \\
 \vac & \occ & \occ \\
\end{array}
\right),
\\[0.3cm]
\left(
\begin{array}{ccc}
 \occ & \occ & \vac \\
 \occ & \vac & \occ \\
\end{array}
\right),
&
\left(
\begin{array}{ccc}
 \occ & \occ & \vac \\
 \occ & \occ & \vac \\
\end{array}
\right),
&
\left(
\begin{array}{ccc}
 \occ & \occ & \occ \\
 \vac & \vac & \occ \\
\end{array}
\right),
&
\left(
\begin{array}{ccc}
 \occ & \occ & \occ \\
 \vac & \occ & \vac \\
\end{array}
\right),
&
\left(
\begin{array}{ccc}
 \occ & \occ & \occ \\
 \occ & \vac & \vac \\
\end{array}
\right)
\end{array}
\right\}.
$}
\end{equation}
The dynamics is defined by moving a particle from some column to a vacant site in the column immediately to the right or the left, if possible. More precisely, the 
following transition,
\begin{equation}
\label{2d-forward}
\begin{array}{cc}
\vdots & \vdots \\
\occ_a & \vdots  \\
\vdots & \vac_b \\
\vdots & \vdots \\
\hline
i & i+1
\end{array}
\longrightarrow
\begin{array}{cc}
\vdots & \vdots \\
\vac & \vdots \\
\vdots  & \occ \\
\vdots & \vdots \\
\hline
i & i+1
\end{array},
\end{equation}
where the $a$'th particle from the \emph{bottom} in column $i$ moves right to the $b$'th vacancy from the \emph{top} in column $i+1$ and there are a total of $k$ particles in column $i+1$, occurs with rate $t^{a+b+k-2}$.
Similarly the reverse transition,
\begin{equation}
\label{2d-reverse}
\begin{array}{cc}
\vdots & \vdots \\
\vdots & \occ_a \\
\vac_b & \vdots \\
\vdots & \vdots \\
\hline
i & i+1
\end{array}
\longrightarrow
\begin{array}{cc}
\vdots & \vdots \\
\vdots & \vac \\
\occ & \vdots \\
\vdots & \vdots \\
\hline
i & i+1
\end{array},
\end{equation}
where the $a$'th particle from the bottom in column $i+1$ moves left to the $b$'th vacancy from the top in column $i$ and there are a total of $k$ particles in column $i$, occurs with rate $q t^{a+b+k-2}$.

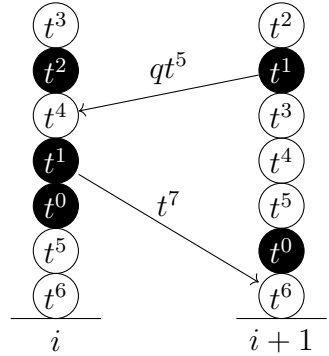
\begin{figure}[h!]
\def\r{3}
\def\sr{0.1*\r}
\begin{center}
\begin{tikzpicture}
    \draw (0,0) circle(\sr cm);
    \node at (0,0*\sr) {\color{black}$t^6$};
    \draw (0,2*\sr) circle(\sr cm);
    \node at (0,2*\sr) {\color{black}$t^5$};
    \draw[fill=black] (0,4*\sr) circle(\sr cm);
    \node at (0,4*\sr) {\color{white}$t^0$};
    \draw[fill=black] (0,6*\sr) circle(\sr cm);
    \node(v1) at (0,6*\sr) {\color{white}$t^1$};
    \draw (0,8*\sr) circle(\sr cm);
    \node(v4) at (0,8*\sr) {\color{black}$t^4$};
    \draw[fill=black] (0,10*\sr) circle(\sr cm);
    \node at (0,10*\sr) {\color{white}$t^2$};
    \draw (0,12*\sr) circle(\sr cm);
    \node at (0,12*\sr) {\color{black}$t^3$};

    \draw (3,0) circle(\sr cm);
    \node(v2) at (3,0*\sr) {\color{black}$t^6$};
    \draw[fill=black] (3,2*\sr) circle(\sr cm);
    \node at (3,2*\sr) {\color{white}$t^0$};
    \draw (3,4*\sr) circle(\sr cm);
    \node at (3,4*\sr) {\color{black}$t^5$};
    \draw (3,6*\sr) circle(\sr cm);
    \node at (3,6*\sr) {\color{black}$t^4$};
    \draw (3,8*\sr) circle(\sr cm);
    \node at (3,8*\sr) {\color{black}$t^3$};
    \draw[fill=black] (3,10*\sr) circle(\sr cm);
    \node(v3) at (3,10*\sr) {\color{white}$t^1$};
    \draw (3,12*\sr) circle(\sr cm);
    \node at (3,12*\sr) {\color{black}$t^2$};

    \tikzstyle{edge} = [->]

    \draw[edge](v1)--(v2);
    \node at (1.5,4.1*\sr) {\color{black}$t^7$};

    \draw[edge](v3)--(v4);
    \node at (1.5,10.1*\sr) {\color{black}$q t^5$};

    \draw (-2*\sr,-\sr) -- (2*\sr,-\sr);
    \node at (0,-2*\sr) {\color{black}$i$};

    \draw (3+-2*\sr,-\sr) -- (3+2*\sr,-\sr);
    \node at (3,-2*\sr) {\color{black}$i+1$};
\end{tikzpicture}

\caption{An example of a forward and backward transition in a system with $K=7$. The forward transition of the second particle from the bottom ($a=2$) at site $i$ to the fifth vacancy from the top ($b=5$) in site $i+1$ with $k = 2$ particles at site $i+1$ according to \eqref{2d-forward} has the transition rate of $t^7$.
Similarly, the backward transition from the second particle from the bottom ($a=2$) at site $i+1$ to the second vacancy from the top ($b=2$) at site $i$ with $k = 3$ particles in site $i$ according to \eqref{2d-reverse} has the transition rate of $qt^5$.}
\label{fig:topbot transitions}
\end{center}
\end{figure}

Since these rates look strange, we motivate them by assigning powers of $t$ to all locations in $\hat{\eta}$ as follows. To the $a$'th particle from the bottom in any column, we assign $t^{a-1}$ and to the $b$'th vacancy from the top in any column, we assign $t^{b+k-1}$ if there are $k$ particles in that column. 
Then the sum of all the $t$-powers in any column is precisely $[K]$. Now, the rate of the forward transition is seen to be exactly the product of the $t$-powers of the particle and the vacancy. The reverse transition has a similar rate with an extra factor of $q$. See \cref{fig:topbot transitions} for an illustration.

These rates are designed so that the steady state probabilities are of product form.
We will now show that the steady state probability of $\hat{\eta} \in \widehat{\Omega}_{L,n}$ is given by
\begin{equation}
\label{2d-statdist}
\pi(\hat{\eta}) = \frac{1}{Z_{L,n}}
\prod_{j=1}^L \prod_{\substack{i=1 \\ \hat{\eta}_{i,j} = \occ}}^{K} t^{i-1},
\end{equation}
where we index the rows from top to bottom.
Note that this formula is not obvious. The rate of a transition depends  only on the relative positions of the particle moving and the vacancy being occupied. But the formula depends on the exact position it occupies.

\subsection{Proof of the steady state formula}
\label{sec:2dss}

It is easy to see that this two-dimensional process is ergodic. Therefore, it is enough to verify that \eqref{2d-statdist} satisfies the master equation.

To prove the formula, we analyze the transitions from three consecutive columns in a configuration $\hat{\eta}$ for now. Suppose the positions of the particles in columns $i-1, i, i+1$ are $\alpha_1, \dots, \alpha_r$, $\beta_1, \dots, \beta_s$ and $\gamma_1, \dots, \gamma_u$ respectively.

\begin{center}
\begin{figure}[h!]
\hspace*{2cm}
\includegraphics[scale=1]{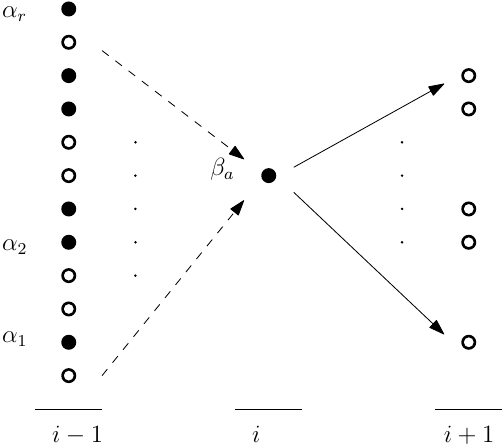}
\caption{Outgoing transitions of the particle at position $\beta_a$ in column $i$ in solid lines and incoming transitions in dashed lines.}
\label{fig:transition}
\end{figure}
\end{center}

Consider all transitions involving the $a$'th particle at position $\beta_a$ in column $i$, as shown in \cref{fig:transition}. 
For the outgoing forward transitions, this particle can move to any of the positions in $\{1, \dots, K\} \setminus \{\gamma_1, \dots, \gamma_u\}$. By \eqref{2d-forward}, the total rate of this transition is
\begin{equation}
\label{2d forward outgoing rate}
t^{a-1+u} (1 + t + \cdots + t^{K-u-1}) 
= t^{a-1} ([K] - [u]).
\end{equation}
Note that if we sum this rate over all particles in column $i$, we obtain \eqref{1d-forward}, which is a good consistency check. The total weight of  forward transitions involving this particle is one term in the $(q, t)$~$K$-ASEP which is $[s]([K]-[u])\pi(\hat{\eta})$.

The incoming forward transitions are a little more tricky to analyze. The particle in column $i$ can arrive from any of the vacancies in column $i-1$, which are at positions $\{1, \dots, K\} \setminus \{\alpha_1, \dots, \alpha_r\}$. For every position between $1$ and $\alpha_1 - 1$, we have a transition according to \eqref{2d-forward} with rate
\begin{equation}
t^{(1) + (K - \beta_a - s + a + 1) + (s-1) - 2} = t^{K - \beta_a + a -1}.
\end{equation}
If the source particle is at position $j$ from the bottom and we call the configuration $\hat{\eta}^{(j)}$, then we get from \eqref{2d-statdist} that
\begin{equation}
\frac{\pi(\hat{\eta}^{(j)})}{\pi(\hat{\eta})} = \frac{t^{K - j}}{t^{K - \beta_a}}
= t^{\beta_a - j}.
\end{equation}
Summing over all possible such positions, the total incoming weight of such transitions is
\begin{equation}
\sum_{j = 1}^{\alpha_1 - 1} t^{K - \beta_a + a -1} \pi(\hat{\eta}^{(j)})
= \sum_{j = 1}^{\alpha_1 - 1} t^{K + a -1 - j} \pi(\hat{\eta})
= t^{K + a} (t^{-2} + \cdots + t^{-\alpha_1}) \pi(\hat{\eta}).
\end{equation}
Similarly, for every position $j$ between $\alpha_1 + 1$ and $\alpha_2 - 1$, the transition rate from the configuration $\hat{\eta}^{(j)}$ to $\eta$ is $t^{K + a - \beta_a}$ and $\frac{\pi(\hat{\eta}^{(j)})}{\pi(\hat{\eta})} = t^{\beta_a - j}$. Thus, the total weight of these transitions is
\begin{equation}
\sum_{j = \alpha_1 + 1}^{\alpha_2 - 1} t^{K - \beta_a + a} \pi(\hat{\eta}^{(j)})
= t^{K + a} (t^{-\alpha_1 - 1} + \cdots + t^{-\alpha_2 + 1}) \pi(\hat{\eta}).
\end{equation}
A similar calculation for every position between $\alpha_2 + 1$ to $\alpha_3 - 1$ leads to the sum
\begin{equation}
t^{K + a} (t^{-\alpha_2} + \cdots + t^{-\alpha_3 + 2}) \pi(\hat{\eta}).
\end{equation}
Continuing this way, we see that when we sum over all positions of vacancies, we will obtain consecutive powers of $t$. Lastly, we need to consider vacancies between positions $\alpha_r + 1$ and $K$. This leads to 
\begin{equation}
t^{K + a} (t^{r - \alpha_r - 2} + \cdots + t^{-K + r - 1}) \pi(\hat{\eta}).
\end{equation}
We now have all incoming forward transitions leading to the particle ending up at position $\beta_a$ in column $i$. Summing all of these gives the grand total
\begin{equation}
\label{2d forward incoming rate}
t^{K + a} (t^{-2} + \cdots + t^{-K + r - 1}) \pi(\hat{\eta})
= t^{a - 1} ([K] - [r])  \pi(\hat{\eta}).
\end{equation}
The important observation is that this rate is independent both of $\beta_a$, as well as the positions $\alpha_1, \dots, \alpha_r$ of particles in column $i-1$. Note also that this quantity is 0 if $r = K$ (in which case $\alpha_i = i$  for $i \in [K]$).
If we now sum over all the particles in column $i$, we will obtain one of the terms in the incoming weights in the $(q, t)$~$K$-ASEP which is $[s]([K]-[r])\pi(\hat{\eta})$. 

{
Similar calculations go through for outgoing and incoming reverse transitions of the particle at position $\beta_a$ in column $i$, the only difference being an extra factor of $q$ in the weights. That is to say, the total outgoing weight in the backward direction is 
\begin{equation}
\label{2d backward outgoing rate}
q \ t^{a-1} ([K] - [r]) \pi(\hat{\eta}),
\end{equation}
and the total incoming rate in the backward direction is
\begin{equation}
\label{2d backward incoming rate}
q \ t^{a - 1} ([K] - [u])  \pi(\hat{\eta}).
\end{equation}
For later purposes, we define a projection $\Pi: \widehat{\Omega}_{L,n} \to \Omega_{L,n}$ given by
\begin{equation}
\label{defproj}
\Pi(\hat{\eta}) = (m_1, \dots, m_L),  \quad
\text{where} \quad m_i= \sum_{j=1}^{K} \hat{\eta}_{i,j}.
\end{equation}
Let $m = \Pi(\hat{\eta})$.
Summing \eqref{2d forward outgoing rate} and \eqref{2d backward outgoing rate} over all the sites and all particles at a site, the total outgoing rate is
\begin{equation}
\label{mastereq-out}
\sum_{i=1}^{L} \, \big( [m_i]([K]-[m_{i+1}]) + q \, [m_i]([K]-[m_{i-1}]) \big) \, \pi(\hat{\eta}).
\end{equation}
Similarly, summing \eqref{2d forward incoming rate} and \eqref{2d backward incoming rate} over all the sites and all particles at a site, the total incoming rate is
\begin{equation}
\label{mastereq-in}
\sum_{i=1}^{L} \, \big( [m_i]([K]-[m_{i-1}]) + q \, [m_i]([K]-[m_{i+1}]) \big) \, \pi(\hat{\eta}).
\end{equation}
To complete the proof, we have to show that \eqref{mastereq-in} and \eqref{mastereq-out} are equal.
These expressions are very similar, except that the factor of $q$ has shifted places. It will suffice to show that
\begin{equation}
    \sum_{i=1}^{L} \, [m_i]([K]-[m_{i+1}]) = \sum_{i=1}^{L} \, [m_i]([K]-[m_{i-1}])
\end{equation}
for every $\hat{\eta}$. Expanding and simplifying, this is equivalent to showing
\begin{equation}
    \sum_{i=1}^{L} \, [m_i][m_{i+1}] = \sum_{i=1}^{L} \, [m_{i-1}][m_i]
\end{equation}
which is obviously true because of periodic boundary conditions. Therefore, we have proved the formula \eqref{2d-statdist} for the steady state probability.
}

\subsection{Proof of projection}
\label{sec:proj}
For a configuration $m \in \Omega_{L,n}$ of the $(q, t)$~$K$-ASEP, we let $\{m \} = \Pi^{-1}(m)$, where the projection $\Pi$ is defined in \eqref{defproj}. Thus, $\{m\}$ is the set of all the states in the two-dimensional model that project to $m$. For the example of $\widehat{\Omega}_{3,4}$ shown in \eqref{2d enriched example}, we have
\begin{equation}
\label{eg-proj}
 \{ (1,1,2) \} = \Bigg\{ \left( \begin{array}{ccc}
 \vac & \vac & \occ \\
 \occ & \occ & \occ \\
\end{array}
\right),
\left( \begin{array}{ccc}
 \vac & \occ & \occ \\
 \occ & \vac & \occ \\
\end{array}
\right),
\left( \begin{array}{ccc}
 \occ & \vac & \occ \\
 \vac & \occ & \occ \\
\end{array}
\right),
\left( \begin{array}{ccc}
 \occ & \occ & \occ \\
 \vac & \vac & \occ \\
\end{array}
\right)
\Bigg\}.
\end{equation}

We define the rate from a two-dimensional state to a one-dimensional state as
\begin{equation}
\label{projrate}
\textrm{rate}(\hat{\eta}\rightarrow m) = \sum_{\hat{\tau}\in\{ m \}} \textrm{rate}(\hat{\eta}\rightarrow \hat{\tau}).
\end{equation}
Notice that $\hat{\eta}$ will not make a transition to all $\hat{\tau} \in \{m\}$. But the rate on the right hand side of \eqref{projrate} will only be nonzero for allowed transitions.
To show that the two-dimensional model projects to the $(q, t)$~$K$-ASEP as a Markov process, we need to prove the \emph{lumping property}~\cite[Lemma~2.5]{levin_peres_wilmer.2009}
\begin{equation}
\textrm{rate}(\hat{\eta}_1 \rightarrow m') = \textrm{rate}(\hat{\eta}_2 \rightarrow m') \quad \forall \, \hat{\eta}_1,\hat{\eta}_2 \in \{m\}, 
\end{equation}
for all $m, m' \in \Omega_{L,n}$. Moreover, we have to show that this rate is the same as $\textrm{rate}(m \rightarrow m')$ for the $(q, t)$~$K$-ASEP model.

We will now prove this result. Let $m, m'$ be states such that a particle jumps from site $i$ in $m$ to $i+1$, namely a forward transition. Then $m'_i = m_i - 1$, $m'_{i+1} = m_i + 1$, and for all $j \neq i, i+1$, $m'_j = m_j$. We have shown that the outgoing rate involving the $a$'th particle in column $i$ in any configuration $\hat{\eta} \in \{m\}$ is given by \eqref{2d forward outgoing rate}, which is independent of $\hat{\eta}$. The sum of rates over all particles in column $i$ is therefore given by \eqref{1d-forward}, which is precisely $\textrm{rate}(m \rightarrow m')$. A similar argument goes through for backward transitions. This completes the proof of projection.

It is a standard result that if a Markov process projects onto another Markov process, then the steady state of the latter can be obtained by summing over the steady state probabilities of the former. In our setting, the steady state weights of both the two-dimensional process and the $(q, t)$~$K$-ASEP are of product form. Therefore, it suffices to look at the weights in one column of the two-dimensional process. 

Consider a single site $j$ in a configuration $\eta$ of the $(q, t)$~$K$-ASEP. The configurations $\hat{\eta}$ which project to it will necessarily have $\eta_j$ particles in the $j$'th column. Thus, the steady state weight of that column is proportional to
\begin{equation}
\wt(\eta_i) = \prod_{\substack{i=1 \\ \hat{\eta}_{i,j} = \occ}}^{K} t^{i-1}.
\end{equation}
We can write the $j$'th column in $\hat{\eta}$ read from bottom to top as a binary word $w$ of length $K$ with $\eta_j$ $1$'s by replacing $\vac \to 0$ and $\occ \to 1$. A \emph{coinversion} in a binary word $w$ is a pair $(a, b)$ such that $a < b$, $w_a = 0$ and $w_b = 1$. The number of coinversions in $w$ is denoted $\coinv(w)$. Then, it is not too difficult to see that $\wt(\eta_i) = t^{\coinv(w) + \eta_i (\eta_i - 1)/2}$. It is a standard result in combinatorics that
\begin{equation}
\sum_{w} t^{\coinv(w)} = \qbinom{K}{\eta_j},
\end{equation}
where the sum runs over all binary words of length $K$ with $\eta_j$ $1$'s~\cite[Proposition~1.7.1]{stanley-ec1}. Therefore, we get that the steady state weight of $\eta$ is given by
\begin{equation}
\wt_\eta(t) \propto \prod_{i = 1}^L t^{\eta_i (\eta_i - 1)/2} 
\qbinom{K}{\eta_i},
\end{equation}
which matches \eqref{wt-eta} up to overall constants.

Continuing \eqref{eg-proj}, $\wt_{(1,1,2)}(t)$ can be obtained using \eqref{2d-statdist} as
\begin{equation}
    \begin{split}
        \wt \left(
\begin{array}{ccc}
 \vac & \vac & \occ \\
 \occ & \occ & \occ \\
\end{array}
\right)= t^3, \quad
& \wt\left(
\begin{array}{ccc}
 \vac & \occ & \occ \\
 \occ & \vac & \occ \\
\end{array}
\right)=  t^2, \\
 \wt\left(
\begin{array}{ccc}
 \occ & \vac & \occ \\
 \vac & \occ & \occ \\
\end{array}
\right)= t^2, \quad
& \wt\left(
\begin{array}{ccc}
 \occ & \occ & \occ \\
 \vac & \vac & \occ \\
\end{array}
\right)= t.\\
    \end{split}
\end{equation}
Therefore, $\wt_{(1,1,2)}(t) = t(1+2t+t^2) = t (1 + t)^2$. To compare with \eqref{ss eg}, note that the sum of weights over all configurations in $\widehat{\Omega}_{3,4}$ will give $3t (t^2+3 t+1)$, which is $t$ times the partition function $Z^2_{3,4}$ given in \eqref{pf eg}.

\section{Conclusions}

In this work, we have focused on a special case of the misanthrope process called the $(q, t)$~$K$-ASEP which is a natural generalization of the ASEP. It is therefore important to study it in its own right.
We have demonstrated several nontrivial properties of the steady state such as particle-hole symmetry and palindromicity. We have also constructed a two-dimensional exclusion process on the cylinder with circumference $L$ and height $K$ which projects to it, and for which the steady steady has nice properties. In particular, the steady state is of product form and is independent of the asymmetry parameter.

It would be an interesting question to investigate how our results can be generalized to formulate a two-dimensional analogue of misanthrope processes with product steady states. In a separate direction, it would also be interesting to explore nonequilibrium properties of the $(q, t)$~$K$-ASEP and see how far the methods for the ASEP, i.e. $K = 1$, extend to this more general framework.

\section*{Acknowledgements}
We thank the anonymous referees for many useful comments.
The first author (AA) thanks Chiara Franceschini for discussions,
and acknowledges support from SERB Core grant CRG/2021/001592 as well as the DST FIST program - 2021 [TPN - 700661].

\bibliography{multiparticle}
\bibliographystyle{alpha}

\appendix

\section{Illustrative example for $\fracpart{n/L} = 1/2$}
\label{sec:eg}

Consider the case of $K = 6$, $L = 4$ and $n \in \{2, 6, 10, 14, 18, 22\}$ so that $\fracpart{n/L} = 1/2$. For each case, we list the numerator of the  probability (as a polynomial in $t$) that $\eta_1$ is equal to $\lfloor n/L \rfloor$ and $\lceil n/L \rceil$ respectively. The claims in \cref{it:1} and \cref{it:2} are seen to work in all these cases.

\begin{itemize}

\item $n = 2$:
\[
\pi(\eta_1 = 0) \propto
3 + 9 t + 9 t^2 + 9 t^3 + 9 t^4 + 9 t^5 + 3 t^6,
\]
and
\[
\pi(\eta_1 = 1) \propto
3 + 6 t + 6 t^2 + 6 t^3 + 6 t^4 + 6 t^5 + 3 t^6.
\]

\item $n = 6$:
\begin{multline*}
\pi(\eta_1 = 1) \propto
3 + 15 t + 54 t^{2} + 135 t^{3} + 303 t^{4} + 585 t^{5} + 1020 t^{6} + 
 1608 t^{7} \\
 + 2346 t^{8} + 3171 t^{9} + 3981 t^{10} + 4677 t^{11} + 5145 t^{12} + 
 5322 t^{13} + 5145 t^{14}  \\
 + 4677 t^{15} + 3981 t^{16} + 3171 t^{17} + 
 2346 t^{18} + 1608 t^{19} + 1020 t^{20} \\
 + 585 t^{21} + 303 t^{22} + 135 t^{23} +  54 t^{24} + 15 t^{25} + 3 t^{26},
\end{multline*}
and
\begin{multline*}
\pi(\eta_1 = 2) \propto
3 + 15 t + 51 t^{2} + 129 t^{3} + 279 t^{4} + 540 t^{5} + 924 t^{6} + 
 1458 t^{7} \\
 + 2100 t^{8} + 2838 t^{9} + 3537 t^{10} + 4161 t^{11} + 4554 t^{12} + 
 4722 t^{13} + 4554 t^{14}  \\
 + 4161 t^{15} + 3537 t^{16} + 2838 t^{17} + 
 2100 t^{18} + 1458 t^{19} + 924 t^{20}\\
  + 540 t^{21} + 279 t^{22} + 129 t^{23} +  51 t^{24} + 15 t^{25} + 3 t^{26}.
\end{multline*}

\item $n = 10$:
\begin{multline*}
\pi(\eta_1 = 2) \propto
3 + 15 t + 57 t^{2} + 165 t^{3} + 405 t^{4} + 876 t^{5} + 1698 t^{6} \\
+ 2994 t^{7} + 4848 t^{8} + 7308 t^{9} + 10230 t^{10} + 13428 t^{11} 
+ 16521 t^{12} \\
+ 19173 t^{13} + 20904 t^{14} + 21540 t^{15} + 20904 t^{16} + 
 19173 t^{17} + 16521 t^{18} \\
 + 13428 t^{19} + 10230 t^{20} + 7308 t^{21} + 
 4848 t^{22} + 2994 t^{23} + 1698 t^{24} \\
 + 876 t^{25} + 405 t^{26} + 165 t^{27} + 
 57 t^{28} + 15 t^{29} + 3 t^{30},
\end{multline*}
and
\begin{multline*}
\pi(\eta_1 = 3) \propto
3 + 15 t + 57 t^{2} + 162 t^{3} + 402 t^{4} + 
 858 t^{5} + 1662 t^{6} \\
 + 2925 t^{7} + 4731 t^{8} + 7095 t^{9} + 9942 t^{10} + 
 13020 t^{11} + 16005 t^{12} \\
 + 18552 t^{13} + 20238 t^{14} + 20826 t^{15} + 
 20238 t^{16} + 18552 t^{17} + 16005 t^{18} \\
 + 13020 t^{19} + 9942 t^{20} + 
 7095 t^{21} + 4731 t^{22} + 2925 t^{23} + 1662 t^{24} \\
 + 858 t^{25} + 402 t^{26} + 162 t^{27} + 57 t^{28} + 15 t^{29} + 3 t^{30}.
\end{multline*}

\item $n = 14$:
\begin{multline*}
\pi(\eta_1 = 3) \propto
3 + 15 t + 57 t^{2} + 162 t^{3} + 402 t^{4} + 
 858 t^{5} + 1662 t^{6} \\
 + 2925 t^{7} + 4731 t^{8} + 7095 t^{9} + 9942 t^{10} + 
 13020 t^{11} + 16005 t^{12} \\
 + 18552 t^{13} + 20238 t^{14} + 20826 t^{15} + 
 20238 t^{16} + 18552 t^{17} + 16005 t^{18} \\
 + 13020 t^{19} + 9942 t^{20} + 
 7095 t^{21} + 4731 t^{22} + 2925 t^{23} + 1662 t^{24} \\
 + 858 t^{25} + 402 t^{26} + 162 t^{27} + 57 t^{28} + 15 t^{29} + 3 t^{30},
\end{multline*}
and
\begin{multline*}
\pi(\eta_1 = 4) \propto
3 + 15 t + 57 t^{2} + 165 t^{3} + 405 t^{4} + 876 t^{5} + 1698 t^{6} \\
+ 2994 t^{7} + 4848 t^{8} + 7308 t^{9} + 10230 t^{10} + 13428 t^{11} 
+ 16521 t^{12} \\
+ 19173 t^{13} + 20904 t^{14} + 21540 t^{15} + 20904 t^{16} + 
 19173 t^{17} + 16521 t^{18} \\
 + 13428 t^{19} + 10230 t^{20} + 7308 t^{21} + 
 4848 t^{22} + 2994 t^{23} + 1698 t^{24} \\
 + 876 t^{25} + 405 t^{26} + 165 t^{27} + 
 57 t^{28} + 15 t^{29} + 3 t^{30}.
\end{multline*}

\item $n = 18$:
\begin{multline*}
\pi(\eta_1 = 4) \propto
3 + 15 t + 51 t^{2} + 129 t^{3} + 279 t^{4} + 540 t^{5} + 924 t^{6} + 
 1458 t^{7} \\
 + 2100 t^{8} + 2838 t^{9} + 3537 t^{10} + 4161 t^{11} + 4554 t^{12} + 
 4722 t^{13} + 4554 t^{14}  \\
 + 4161 t^{15} + 3537 t^{16} + 2838 t^{17} + 
 2100 t^{18} + 1458 t^{19} + 924 t^{20}\\
  + 540 t^{21} + 279 t^{22} + 129 t^{23} +  51 t^{24} + 15 t^{25} + 3 t^{26},
\end{multline*}
and
\begin{multline*}
\pi(\eta_1 = 5) \propto
3 + 15 t + 54 t^{2} + 135 t^{3} + 303 t^{4} + 585 t^{5} + 1020 t^{6} + 
 1608 t^{7} \\
 + 2346 t^{8} + 3171 t^{9} + 3981 t^{10} + 4677 t^{11} + 5145 t^{12} + 
 5322 t^{13} + 5145 t^{14}  \\
 + 4677 t^{15} + 3981 t^{16} + 3171 t^{17} + 
 2346 t^{18} + 1608 t^{19} + 1020 t^{20} \\
 + 585 t^{21} + 303 t^{22} + 135 t^{23} +  54 t^{24} + 15 t^{25} + 3 t^{26}.
\end{multline*}

\item $n = 22$:
\[
\pi(\eta_1 = 5) \propto
3 + 6 t + 6 t^2 + 6 t^3 + 6 t^4 + 6 t^5 + 3 t^6,
\]
and
\[
\pi(\eta_1 = 6) \propto
3 + 9 t + 9 t^2 + 9 t^3 + 9 t^4 + 9 t^5 + 3 t^6.
\]

\end{itemize}

\end{document}